# Experimental, ad hoc, online, inter-university student e-contest during the pandemic – Lessons learned

Horia-Nicolai Teodorescu [a,b]
[a] Lab. BME, 'Gh. Asachi' Technical University of Iasi
[b] Institute of Computer Science of the Romanian Academy
Iasi, Romania
hteodor@etti.tuiasi.ro

*Abstract*—We are reporting on lessons learned from an e-contest for students held during the current pandemic. We compare the e-contest with the 10 previous editions of the same, but face-to-face contest. While apparently the competition did not suffer because of being a virtual one, some disadvantages were noted. The main conclusions are: the basic interconnectivity means arise no serious technical issue, but the jury-jury interconnectivity is more limited than the face-to-face one; online jury-competitors interactivity is poorer than face-to-face interactivity; human factors, higher uncertainties in the organization process, and less time to spend in the process for the local organizers are major limiting factors; concerns on the participation and evaluation fairness are higher; involuntary gender discrimination seems lower, but persists; there are serious concerns related to privacy, including differential privacy; some peculiarities of the presented topics and of the evaluation process emerged, but it is unclear if they are related to the online nature of the competition, to the extra stress on the particiapants during the pandemic, or to other factors, or are random. While some conclusions may be intimately related to the analyzed case, some are general enough for being worth to other online competitions and examinations.

*Keywords—online education, online examination, student competition, pandemics, fairness, privacy, differential privacy, education experiment, microsystems.*

## I. Introduction

Numerous studies have been recently devoted to the need of and requirements for online learning under the conditions created by the current pandemic [1-6]. However, no such study was published on extra-curricular activities, such as student competitions. The topic becomes more important as several countries plan online national-wide examinations (A-levels, GCSEs) and vocational qualifications, and universities struggle to find methods of replacing various types of examinations, including graduating examinations with online ones or surrogates, such as 'prediction of results' in UK and Ireland [7] or with other "grades issued this summer by the relevant recognised authority" as Oxford University vaguely describes [8].

Online competitions are not new – such competitions are regularly organized by universities in the field of computer science, in Romania and in many countries. In the domain of hardware, probably the best known international competition is the Create the Future Design Contest, which "was launched in 2002 by the publishers of NASA Tech Briefs magazine to help stimulate and reward engineering innovation" [9]. Also, research project competitions are the rule in EU, USA, and many other countries. However, none of these took place in conditions similar to those of the current pandemic, and only few have been thoroughly analyzed and publically documented.

We report on an experimental, online, inter-university student project competition that was held on May 15, 2020, under the lockdown conditions of the current pandemic. Although this edition was intently planned to be experimental and it cannot be considered a regular competition, in fact it was the eleventh organized by us on the same topic under the title "'M. Konteschweller' Competition on Microcontrollers and Microsystems" (since 2009). Details on the competition are given in the Annex.

Because the competition we organized and report on was one of the very few to have been held before May 15, 2020, at least as far as we are aware, we report on it with the hope that the lessons learned from this competition can help improving future ones. While many of the reported facts are anecdotic because they lack the large statistic background for validation, they may be indicative for further research directions.

## II. Participants and Organization Facts

This is a student project competition with the only restriction related to the topic of the project, which must include at least one operational microcontroller or microsystem. The projects are judged by their complexity, innovative ideas, applicability, level of functioning and operability, presentation, and demonstrated knowledge of the presenting students. The open theme of the competition differentiates it from the similar contest on designing with microcontrollers, organized by the firm Continental Iasi, which is also annually held, but has a single annual theme determined by the firm.

The participating universities and the number of teams and participants from each university are shown in Table 1. The students represented departments of electronics, communications, and computer engineering. Students were enrolled in the second, third, fourth, and fifth study years. Companies supporting the competition, offering their own prizes to students, and having members in the jury included, during the years, Continental, Infineon, Microchip, MobileService, Preh, Veoneer, and Silicon Service [10], [11] and Mechatronics. The Ministry of Education also provided financial support for accommodation and meals, during the contest, to the students from other cities.

TABLE I. PARTICIPATING UNIVERSITIES IN THE EXPERIMENTAL, ONLINE STUDENT COMPETITION DURING THE LOCKDOWN

| University (U) and city | Number of teams | Number of students |
|---|---|---|
| 'Politehnica' TU[a] Bucharest | 3 | 5 |
| TU Cluj-Napoca (Romania) | 3 | 3 |
| TU of R, Moldova, Chisinau | 2 | 3 |
| U. 'Stefan cel Mare' Suceava (Romania) | 3 | 6 |
| U. 'Dunarea de Jos' Galati (Romania) | 3 | 6 |
| TU 'Gheorghe Asachi' Iasi (Romania) | 4 | 7 |

[a.] TU stands for Technical University

Presentations were accepted for registration up to the very morning of the competition. The average time of presentation was similar with that for previous, face-to-face editions of the contest. A comparison of the 2020 contest with the previous ones in its series is given in Table 2.

TABLE II. COMPARISON BETWEEN THE EXPERIMENTAL 2020 EDITION AND THE PREVIOUS 10 EDITIONS OF THE CONTEST

| Feature | 2020 (during pandemic) | Previous[a] 10 contests (average) |
|---|---|---|
| Number of student teams | 18 | 24 |
| Number of students per team | 1-3 | 1-4 |
| Number of participating universities | 6 | 6 |
| Number of members of the jury | 6 | 12 |
| Ratio of academic vs. firms members of the jury | 4/3 | 1/1 |
| Number of supporting firms | 4 | 7 |
| Number of local organizers active in the online contest | 2 | 4 |
| Time for students teams to prepare (months) | 3 | 5 |

[a.] Based on incomplete personal archive

III. INSURING THE QUALITY OF THE PROJECT EVALUATION PROCESS: ACADEMIA-INDUSTRY JURY

The uniformity of grading was a serious concern because of the inherently limited interactivity of the jury with the teams, offline, because of the lack of hands-on interaction of the jury with the presented projects, and because of the limited or null interactivity of the members of the jury during the presentations – unlike face-to-face competitions. Therefore, this crucial matter is investigated first.

The uniformity of grading is a good test for the homogeneity of purposes during grading, that is, clarity of grading criteria. The standard deviation of the grades per member of the jury is also a good indicator of efficient evaluation; the dispersions of grades assigned by the members of the jury are similar, denoting consistency in grading.

At the team level, the standard deviation of the evaluation by the members of the jury is ideally null; however, this may not be good for the evaluation process, because each member of the jury may have specific competencies. Interestingly, null STDEV was obtained for the lowest grade and for the third and fourth highest (both graded with 9.00).

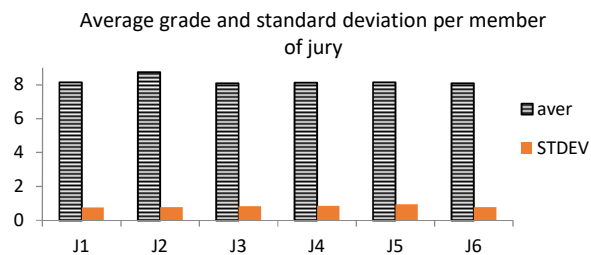

Fig. 1. The grading by the members of the jury (J1 to J6 represent individual jury members).

In principle, there should be no relationship between the spreading and the average of the grades of the teams. The average grade per contest was 8.22 and the STDEV of all grades was 0.68. The average of the STDEV of the individual grades given by the jury members, for all the teams was 0.52, which shows a reasonable uniformity in the grading, only slightly larger than the random choice between successive grades (which is 0.5).

Surprisingly, we found that there is a weak correlation between the overall grade (average grade) and the standard deviation of the grades assigned by the jury for the team, see Fig. 2. The slope of the regression is negative (-0.16) meaning that for every extra point in the grade, the precision of the grade is 0.16 point better ($R^2=0.115$). There are three 'outlier' grades, with zero standard deviations. If the two outliers are removed, the slope becomes -0.2 and the coefficient of determination of the linear regression is even higher ($R^2=0.29$), Fig. 3. This result call into question the grading process improvement, with more time spent with those teams less convincing.

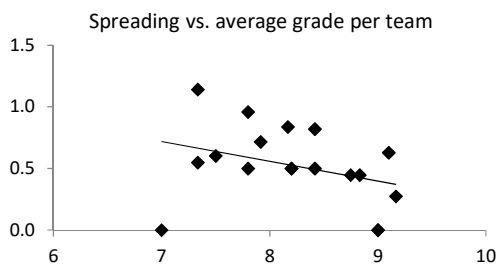

Fig. 2. The average grade of each team vs. the standard deviation of the grades assigned by the members of the jury to the team.

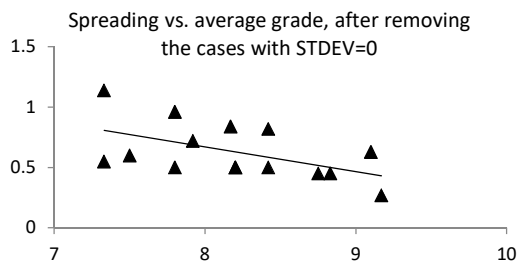

Fig. 3. Regression after removing the outliers

IV. TECHNICAL LIMITATIONS AND CONCERNS RELATED TO THE ONLINE PARTICIPATION AND EVALUATION

The practice of online classes, examinations [12], and of the reported contest indicate that, despite the technical progresses, the platforms for audio-video interaction over the Internet offer a much more limited degree of freedom in the interaction. The difficulties of fast and easily zooming in and out during practical demonstrations and of fast, eye-blink change of the direction of view were evident from the student presentations. The inability for the jury members to privately or semi-privately consult and discuss with each other, as in a face-to-face conversation, without disturbing the other members of the jury or stopping the presentation is annoying in online interactions and limits the efficiency of the evaluation process. The good, yet limited quality of the online pictures was challenging when details of the devices were needed. Overall, the online paralinguistic communication is still poor and at time frustrating in the educational process, including in the online competitions. Unfortunately, these issues cannot be solved before the summer and autumn examinations planned in 2020, under the COVID-19 pandemic restrictions.

The online presentation in the contest was found to have several disadvantages:

- Lack of interactivity of the students; the students don't make friends, don't exchange impressions about their campuses, universities, teachers, employability and student life;
- Students don't interact easily and fully with their colleagues and with the jury;
- Although we have not a proper statistic, it seems that the average number of questions asked by the members of the jury per project was slightly lower than the same number during the previous, face-to-face competitions.

On the other hand, as already said, the online evaluation posed no problem, with low overall average and maximal spreading between the grades between the jury members, for all the participants (Fig. 1).

## V. Fairness and Discrimination Concerns

Discrimination was also a concern in this experimental competition. Fairness in grading and prize allocation is always an issue when students are enrolled in different study years. Among the many possible approaches to fairness, in all the eleven editions of this contest, the study-year blind approach was used. This fairness criterion is considered, in fact, unfair. Small adjustments have been made, including this year, by offering a honorable mention diploma to the youngest participant(s). However, this approach is disputable and we plan to adjust the grades by coefficients determined on a larger statistics of the grades in previous years.

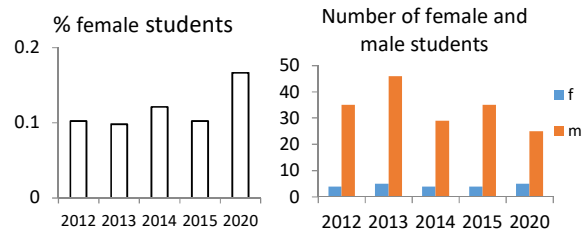

Fig. 4. Percentage of female participants and male vs. female participants.

The average composition in electronics, communications, information technology and computer engineering in the Romanian universities is about 40% female and 60% male students. In the contest discussed here, the percentage of female students was only 17% (5 female vs. 25 male students), which is problematic. Yet, the percentage was larger than in previous years (2012-2015). From this point of view, the competition was unfair during all years and the situation has to be corrected in the future. We conclude that involuntary gender discrimination in the student participation seems to have been lower in this contest during the pandemic, a phenomenon opposed to the one noticed by other authors in other circumstances [13-17]. This finding asks for correcting actions in such contests.

Of special concern is that only one member of the jury was a woman. The situation was the same for several other editions of the competition, and in many previous editions the jury was all-male. This may discourage female participants. The issue needs to be corrected in future editions and should be considered in all online contests, because that may have even more impact than in face-to-face competitions.

## VI. Lexical and Semantic Analysis of the Contest

It may seem that the NLP-type analysis of the presentation is useless. However, similar studies in various domains, from marketing to patent analysis, have proved able to reveal interesting aspects based on limited information from the documents, for example based on similarities of patent titles, abstracts, and claims [18,19]. Also, NLP analysis is well known to reveal stress and sentiments. Because the pandemic, the lockdown, and the novelty of the online competition, stress was probable. Therefore, an NLP analysis is justified.

The titles and textual content of the presentations have not been analyzed in relation with student contests, we believe. Yet, such an analysis, even when reduced to the titles of the presentations may shed some light on the level of the presentations. The titles of the projects had an average number of words of 7.45, and a standard deviation of 5.6 words; the number of words in the titles varied between 3 and 21. The three longer titles were 21, 18 and 17 words length. The average length of the words in the titles was 6.5; the total number of words in the titles was 134 and the total number of letters (without spaces) was 866.

The number of lemmas used in the titles is 77; 68 of them occur once. Nine lemmas are repeated twice or several times. These repetitions are instructive, because they show how many projects may adopt the same broad pint of view, or have similar topics, or use similar approaches.

TABLE III.

| Single occurrence lemmas | Total number of words (vocabulary) | Number of lemmas |
|---|---|---|
| 68 | 97 | 77 |

We use two indices of coherence of the topics [12] of the competition as follows:

$$I_{C1} = \frac{count\ of\ number\ of\ repeated\ words}{number\ of\ unique\ words} = 0.43 \quad (1)$$

This index indicates how much the titles are innovative and consequently how much the titles are similar at the lexical level because of repeated words. When computing $I_{C1}$, stop words are removed.

$$I_{C2} = \frac{number\ of\ titles\ with\ double\ occurence\ words}{total\ number\ of\ titles} = 0.435 \quad (2)$$

This index indicates the degree of similarity between the titles of the presentations in the competition. When computing $I_{C1}$, stop words are preserved. Notice that $I_{C2}$ also includes titles with more than a single word in common.

Both indices are less than but not far from 0.5, thus indicated a moderate spreading of the topics of the projects. The indices may inform project advisors on the need to be more innovative and on the trends in the interests of the students.

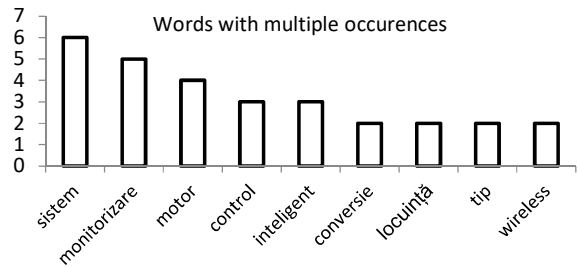

Fig. 5. Most frequent words in the title of the presentation (in the experimental, online edition of the competition, in 2020).

The most frequent words (Fig. 5) indicate that participants emphasized on the complexity of the projects (using the words "system" and "intelligent") and addressed mainly monitoring ("monitorizare") and control applications, followed by energy conversion. The rank-frequency distribution of words, after removing the stop-words (which is not usually done in this kind of analysis), does not follow Zipf's law, but this may be due to the very low number of cases in the statistic.

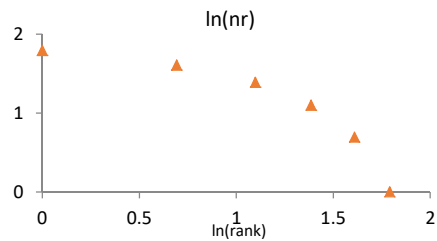

Fig. 6. Number of lemmas vs. rank words in the titles of the presentation, in the 2020 competition

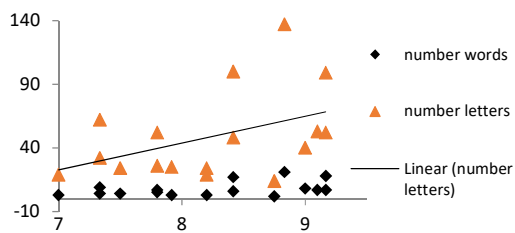

Fig. 7. Apparent relationship between the number of words in the titles of the presentation and the grade, in the 2020 competition.

A fact that is surprising is that 6 of the first 7 projects had a longer title than the average, see Fig. 7. The linear regression of number of letters in the title vs. grade has a positive slope and the coefficient of determination is 0.187. This may be due to the greater care for details in the best projects, or may be a random occurrence. This is however not a rule valid for former editions of the competition, for example in 2018 the first prize was won by a project with title composed of only 29 letters, the second had 33 letters, and the third 25 letters (see [12]). On the other hand, there were long titles in previous competitions too, but they were less frequent. (Examples of longer titles in 2012-2018 competitions are 'Sistem embedded auto de comanda si control a ferestrelor cu protectia utilizatorului si a sistemului electric', with 16 words and 94 characters, 'Robot inteligent care detecteaza succesiuni de marcaje și îndeplineste task-uri, simulând un oraș inteligent' with 14 words and 95 characters,and 'Detectie culoare /obiect prin intermediul microcontrolerului folosind mediul de programare Matlab', with 11 words and 87 characters.) Therefore, the question of usefulness of the NLP analysis in detecting issues in online examinations and competitions remains an open one.

VII. Privacy, Limited Differential Privacy, and Security Concerns – Online Competitions

*Privacy issues*

Competitions raise significant privacy issues for the participants. While student wish to participate to gain recognition and increase their chance of early employability, they may also fear that a poor performance may harm their future employability and peers' respect. When the results of the competition are kept in company databases or are made openly public, the harm peril is very real, as is the perspective of gains for the best performers. The amount of data allowed to be stored in competitions is an issue. When only very limited data is recorded, such as the name of the participants and the prizes they obtain, if any, the dangers are much lower than when full video recordings are made, which may be used in human resources departments in various ways. Openly posting on the Internet the contest may bring much harm to all participants and the jury as well. Therefore, the way of documenting the contest should be chosen with great care and with a legal perspective too.

Pandemics pose new challenges to privacy and security [20], [21]. As any limited-public and partly public-restricted performance, an Internet based competition may pose certain privacy issues. For examples, the student presentations are *not* public in a face-to-face contest as long as the student teams do not choose to make public themselves. Also, the members of the jury and the participants are not meant to be filmed and then shown in live movies in a face-to-face contest, except when they agree so in written form and after being suitably informed. In this way, a limited degree of openness – and a limited degree of privacy – according to the "contained public venue" vs. "open public venue (see [22]) ", or equivalent "semi-open learning environments" [23] was enforced.

Even when a written consent is obtained, the person posting has responsibilities in the use of the content; moreover, copyright issues may occur with the content of the presentations. Therefore, we chose to clearly inform, from the beginning of the competition, all the participants and the members of the jury that taking pictures or movies is not allowed, except pictures of themselves. However, we still feel that the legal background for student competitions on the Internet is unclear and should be addressed in the near future, if the Internet based education were to expand. For example, the competition organizers have no means to control that the participants do not take pictures and to enforce several of the competition rules. There is no clear reference in the legislation on the responsibilities of the participants and of the organizers – a gap that should be filled in as soon as possible. Actually, the same deficiencies apply to online classes and examinations.

*Limited differential privacy issues*

The differential privacy problem is to provide complete information on a population while preserving full anonymity of the individuals. The problem is key to learning because the learner needs feedback from the teacher and vice versa, moreover improvements of the learning process require detailed knowledge on the learner population. The problem is known to have no solution; therefore, limited differential privacy is the main topic of research [24], [25]. Adapting in the case of a contest the words of [24], the differential privacy dilemma for participants is to decide on the "difference between the probability of harm given that they participate and the probability of harm given that they do not participate" in the contest, taking into account the ways the contest is documented and its level of public openness.

For understanding the differential privacy issue in the context of competitions where the competitors know who the members of the jury are, and vice versa, consider an example. Because firms bestow their own prizes, the specific firm prize winning competitor easily derives that the firm representative in the jury gave her a large grade, while the other firm representatives have not conferred the largest possible grade. Thus, the competitors gain some very specific information at least on some of the members of the jury. When a competitor gets two firm awards out of the possible few, but does not win any general prize of the competition (for which the academic members of the jury have a vote), information on the vote of the professors is obtained. That information, in turn, may influence the grading of a professor by the respective student. To remove this and fairness concerns, the academic members of the jury should not be allowed to vote for students from their university. Also, it is advisable that the prizes offered by the firms are proposed as sets of choices by the firm representatives and decided with the vote of the academic members of the jury. This would reduce the information gained from the firm prizes.

Under crises, there is a danger that people have less time to decide on privacy issue and that the competitions are more publically open that they need to be. Also, the online character, even when with a limited public, increases the odds of privacy issues. One member of the jury did not agree to use video communication, using audio only. Also, concerns for the members of the jury may have arisen because his supervisors were also able to see the jury deliberations, although he was not a participant in the jury,

*Security issues*

The security concerns relate to insuring the security and non-disclosure of the files of the deliberations of the jury, the private and personal data of the jury members, and the private data of the participants, including e-mail addresses, grades obtained, and potentially sensitive details of their projects. The security measures the organizing team of the competition has at hand are scarce, if any, beyond those in place at their Internet provider. Especially in times of crises, legal protection for the organizers, members of the jury, participants, and universities have to be established more clearly. As far as we were able to determine, such legal protections are unavailable today in any country in EU and North America.

VIII. CONCLUSIONS

There are several conclusions that may be useful for the organization of other online student competition under harsh conditions. The main positive conclusion is that the interconnectivity means arise no serious technical issues; current broadband connectivity and numerous good platforms allow reliable and good enough, although limited interactivity. However, it was apparent that online jury-competitors interactivity is poorer than face-to-face interactivity and may have created some frustrations on both sides. The main issues are related to human factors, to higher uncertainties in the organization process, and to much shorter time to spend in the process by the people in the local organizing team. There are several issues specific to the online contest, including concerns on the lower degree of fairness in the participation and evaluation, and serious concerns related to privacy. The analysis of the evaluation process and of the topics of the presentations revealed some unexpected peculiarities, but it is not clear if these are due to the online nature of the competition, to other factors, or are random variations.

**Disclaimer.** The article reflects only the opinions of the author.

**Conflict of interests.** HNT is the initiator of the contest in 2009 and has been a member of the organizing team since then.

ACKNOWLEDGMENT

I thank Prof. M. Hagan, the main organizer of the reported online competition (2020 edition of contest 'M. Konteschweller), Prof. C. Aghion and Prof. M. Zbancioc, who have been members of the organizing committee of all the editions (in 2020, up to the pandemic incept), and to Prof. D. Tărniceriu, Dean of the local organizing Faculty, for constant support. I acknowledge that several ideas on ethical issues, privacy, security and NLP analysis originated from Prof. M. Teodorescu, Boston, who also indicated several references. I acknowledge the support of the TU of Iasi and of the Ministry of Education for the first 10 editions of this student competition. I apologize to the members of the jury and to all participants for any inconvenience during this online competition and I ensure them that, if any, such inconvenience was totally involuntary.

REFERENCES

[1] O. Peterson and A. Thankom, Spillover of COVID-19: Impact on the global economy. 2020 Online at https://mpra.ub.uni-muenchen.de/99850/ MPRA Paper No. 99850, posted 26 Apr 2020. Accessed May 20, 2020.

[2] M. Javaid A. Haleem, Vaishya R, Bahl S, Suman R, Vaish A. Industry 4.0 technologies and their applications in fighting COVID-19 pandemic [published online ahead of print, 2020 Apr 24]. Diabetes Metab Syndr. 2020;14(4):419-422. doi:10.1016/j.dsx.2020.04.032

[3] C. Mae Toquero, Challenges and Opportunities for Higher Education amid the COVID-19 Pandemic: The Philippine Context. Pedagogical Research 2020, 5(4), em0063e-ISSN:2468-4929 https://www.pedagogicalresearch.comReview. Accessed May 20, 2020.

[4] S.L. Schneider, Council ML. Distance learning in the era of COVID-19, published online ahead of print, 2020 May 8. *Arch Dermatol Res*. 2020;1-2. doi:10.1007/s00403-020-02088-9

[5] G. Wang et al., Mitigate the effects of home confinement on children during the COVID-19 outbreak. The Lancet, Vol. 395, Issue 10228, 945 - 947.

[6] G. Chowell et al., The COVID-19 pandemic in the USA: what might we expect? The Lancet, Vol. 395, Issue 10230, 1093 – 1094.

[7] S. Harrison, Coronavirus: Predicted grades replace Irish Leaving Cert exam. https://www.bbc.com/news/world-europe-52592956. Accessed May 22, 2020.

[8] Oxford University, Coronavirus (COVID-19): advice for applicants and offer holders, https://www.ox.ac.uk/students/coronavirus-advice/offer-holders-and-applicants?wssl=1. Accessed May 21, 2020.

[9] Create the Future Design Contest, https://contest.techbriefs.com/about. Accessed May 21, 2020.

[10] "Proiecte îndrăznețe și inovatoare ale studenților, prezentate la Concursul „Mihail Konteschweller" găzduit de Universitatea Tehnică". 19 aprilie 2018. https://www.tuiasi.ro/noutati/proiecte-indraznete-si-inovatoare-ale-studentilor-prezentate-la-concursul-mihail-konteschweller-gazduit-de-universitatea-tehnica/. Accessed May 21, 2020.

[11] 60 de studenti din 7 universitati in concurs la TU Iasi. Ieseanul (newspaper), 11.05.2015, https://www.ieseanul.com/60-de-studenti-din-7-universitati-in-concurs-la-tuiasi/. Accessed May 21, 2020.

[12] M. Teodorescu, Personal communication, 2020.

[13] Y. Awwad, R. Fletcher, D. Frey, A. Gandhi, M. Najafian, M. Teodorescu, 2020. Exploring fairness in Machine Learning for international development. MIT D-Lab, Report. Cambridge: MIT D-Lab. https://d-lab.mit.edu/sites/default/files/inline-files/Exploring_fairness_in_machine_learning_for_international_development_03242020_pages_0.pdf. Also, Fairness, bias, and appropriate use of Machine Learning. MIT. https://d-lab.mit.edu/research/mit-d-lab-cite/fairness-bias-and-appropriate-use-machine-learning. Accessed May 3, 2020.

[14] J.J.V. Bavel, Baicker, K., Boggio, P.S. et al., Using social and behavioural science to support COVID-19 pandemic response. Nat Hum Behav 4, 460–471 (2020). https://doi.org/10.1038/s41562-020-0884-z.

[15] WHO Team, A guide to preventing and addressing social stigma associated with COVID-19. 24 Feb 2020. https://www.who.int/who-documents-detail/a-guide-to-preventing-and-addressing-social-stigma-associated-with-covid-19?gclid=EAIaIQobChMIs8-VqtfG6QIV1u5RCh3W_gcfEAAYASAAEgI0fvD_BwE. Accessed May 19, 2020.

[16] A. Krosch, The pandemic could lead to more discrimination against black people. Scientific American, Apr 23, 2020. https://blogs.scientificamerican.com/voices/the-pandemic-could-lead-to-more-discrimination-against-black-people/. Accessed May 21, 2020.

[17] N. Mawar, Saha S, Pandit A, Mahajan U., The third phase of HIV pandemic: social consequences of HIV/AIDS stigma & discrimination & future needs. Indian J Med Res. 2005;122 (6):471-484.

[18] C. deGrazia, N. Pairolero, M. Teodorescu, Shorter patent pendency without sacrificing quality: The use of examiner's amendments at the USPTO (June 2019). USPTO Economic Working Paper No. 2019-03. Available at SSRN: https://ssrn.com/abstract=3416891.

[19] M. Teodorescu, Machine Learning methods for strategy research. Harvard Business School Research Paper Series, Report number 18-011. Publication date 2017/7/31

[20] B. McCall, Shut down and reboot—preparing to minimise infection in a post-COVID-19 era. The Lancet Digital Health. Apr 28, 2020. DOI:https://doi.org/10.1016/S2589-7500(20)30103-5. Accessed May 21, 2020.

[21] S. Vaudenay, Centralized or decentralized? The contact tracing dilemma. https://eprint.iacr.org/2020/531.pdf, 2020, May 6[th].

[22] Patrick R. Lowenthal, David Thomas, The death of the digital dropbox: Rethinking student privacy and public performance. In: The CU Online Handbook, 2011, P.R. Lowenthal, D. Thomas, B. Yuhnke, A. Thai, M. Edwards, C. Gasell (Eds.). https://works.bepress.com. Accessed May 21, 2020.

[23] B. Lorenz, S. Sousa, and V. Tomberg, Privacy awareness of students and its impact on online learning participation – A Case Study. In T. Ley et al. (Eds.): OST 2012, IFIP AICT 395, pp. 189–192, 2013.

[24] C. Dwork, A. Roth, The algorithmic foundations of differential privacy. Foundations and Trends in Theoretical Computer Science, Vol. 9, Nos. 3–4 (2014) 211–407.

[25] N. Agarwal, K. Singh, The price of differential privacy for online learning, Proc. The 34th Int. Conf. Machine Learning - ICML'17, Vol. 70, pp. 32–40, Sydney, NSW, Australia, 2017, JMLR.org.

**Annex: The Student Competition of Microsystems and Microcontrollers 'M. Konteschweller', 2009-2020**

The competition is related to micro-systems and microcontrollers and their applications. Representatives of universities and firms in the field always participated in the contest jury. This year was the eleventh of the contest and it was the first to be organized fully online.

M. Konteschweller was a Romanian-born engineer who worked at Bristol, U.K., France, and Romania; he demonstrated radio remote control of 'robotic' devices such as boats in the early 1930's in Romania; he also served as a professor in the Polytechnic (technical) university of Iasi. He was an inventor with the patent GB400467A "Improvements in or relating to motor cars" obtained in U.K. in 1933 (1933-01-25) [A1], cited in [A2].

According to an account of the TU Gheorghe Asachi of Iasi, [10 "Proiecte], "This is one of the few competitions without an imposed theme", and "A good feature of this competition is that we attract and encourage two … categories: on one side the students on the other the companies, and we put them face-to-face." (Prof. C. Aghion, in [11]). This has as a result that the students, "irrespective of the prize, all are winners" (Dean of the local organizing Faculty, Prof. Daniela Tărniceriu) [10].

[A1] GB400467A, priority 1933-01-25, published 1933-10-26, Mihai Konteschweller, Improvements in or relating to motor cars.

[A2] US2822214A, Interconnected vehicle front door and steering wheel for easy access to seats. 1953.